\documentclass[11pt,a4paper]{article}
\usepackage{times,latexsym}
\usepackage{url}
\usepackage[T1]{fontenc}

\usepackage[acceptedWithA]{tacl2021v1}
\usepackage{xspace,mfirstuc,tabulary}

\usepackage{amsmath}
\usepackage{enumitem}
\usepackage{booktabs}
\usepackage{lscape} 
\usepackage{graphicx}
\usepackage{subcaption}
\usepackage{xspace}
\usepackage{arydshln}
\usepackage{bbm}
\usepackage[normalem]{ulem}
\usepackage{multirow}
\usepackage{array,tabularx}
\usepackage{amssymb}
\usepackage[acronym]{glossaries}
\usepackage{makecell}
\usepackage{textcomp}
\usepackage{rotating,amsmath,booktabs}
\usepackage{cleveref}
\usepackage{enumitem}
\usepackage{soul}
\usepackage{tcolorbox}

\usepackage{booktabs}
\usepackage{multirow}
\usepackage{hyperref}

\usepackage{algorithm}
\usepackage{algpseudocode}
\usepackage{soul}

\newif\iftaclinstructions
\taclinstructionsfalse % AUTHORS: do NOT set this to true
\iftaclinstructions
\renewcommand{\confidential}{}
\renewcommand{\anonsubtext}{(No author info supplied here, for consistency with
TACL-submission anonymization requirements)}
\newcommand{\instr}
\fi

\iftaclpubformat % this "if" is set by the choice of options

\else

\fi

\newtcolorbox{keywords}{colback=yellow!5!white,colframe=yellow!75!black,text width=0.95\linewidth,fontupper=\small,top=3pt,bottom=2pt,left=2pt,right=2pt}

\newtcolorbox{finding}{colback=blue!5!white,colframe=blue!35!white,text width=0.95\linewidth,fontupper=\normalsize,top=2pt,bottom=1pt,left=2pt,right=2pt}

\newtcolorbox{recommendation}{colback=red!5!white,colframe=red!75!black,text width=0.96\linewidth,fontupper=\normalsize,top=3pt,bottom=2pt,left=2pt,right=2pt}

\title{Structuring the Space of Sociotechnical Alignment: A Specification Framework and Systematic Literature Review}

\author{Esra Dönmez and Agnieszka Falenska \\
    Institute for Natural Language Processing, University of Stuttgart\\
    Interchange Forum for Reflecting on Intelligent Systems, University of Stuttgart\\
    \normalsize{\texttt{esra.doenmez@ims.uni-stuttgart.de}}}

\date{}

\begin{document}
\maketitle
\begin{abstract}
Sociotechnical alignment concerns the \emph{social desirability of AI behavior} and is thus inherently \textbf{\emph{normative}}, not merely technical. While NLP research increasingly addresses its technical aspects, it often leaves underspecified what such ``social desirability'' entails. We argue that this reflects a fundamental gap: the absence of a systematic way to specify how sociotechnical alignment defines, justifies, and evaluates socially desirable AI behavior.

To address this gap, we introduce a human-centered framework for specifying sociotechnical alignment.
We draw on social-scientific accounts of sociobehavioral desirability to ground the basis for behavioral desirability judgments and use this framework to analyze how alignment is specified in practice. Our systematic literature review identifies recurring patterns: normative concepts grounding desirability judgments are often unspecified or conflated with alignment targets for (desired) system behavior, target populations are underdefined, and design choices are rarely theoretically justified. These findings point to a lack of conceptual specificity that limits cumulative progress. 
We therefore offer recommendations that link social-scientific frameworks to alignment design choices, supporting more conceptually precise approaches to sociotechnical alignment.
\end{abstract}

\section{Introduction}
As AI systems become embedded in social contexts, they must meet not only functional goals, such as producing accurate recommendations, but also \emph{normative expectations} about how they should \emph{``behave''} in society, for example, by being fair and respectful in their interactions \cite{Selbst2019Fairness,weidinger-etal-2024-star}. This broader challenge of specifying, modeling, and evaluating socially desirable AI behavior across social contexts is referred to as \emph{sociotechnical alignment}.

\begin{figure}[t]
    \centering
    \includegraphics[width=1\linewidth]{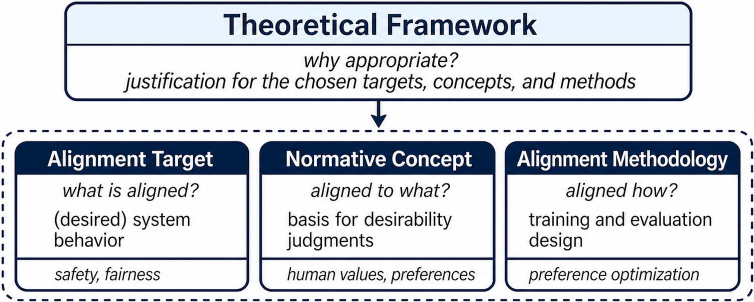}
    \caption{Framework for specifying sociotechnical alignment: overarching theoretical framework at the top, with three dimensions below: alignment target, normative concept, and alignment methodology, each with examples at the bottom.}
    \label{fig:framework}
\end{figure}

Despite growing work on sociotechnical alignment, the literature still leaves ``social desirability'' underspecified. First, ``alignment'' is often used as a catch-all term for heterogeneous, socially contested objectives, ranging from user engagement to social acceptability \cite{Gabriel2023TheChallenge}. Second, existing work often moves between different ways of specifying what AI systems should ``align to’’, ranging from normative concepts (e.g., human values) to more instrumental proxies (e.g., user preferences or stakeholder requirements), without specifying the reasoning behind these choices \cite{Ji2025AIAlignment}. This ambiguity is worsened when alignment is framed primarily in terms of ``human values'', blurring technical questions of system behavior with normative disputes over whose values matter and how they are interpreted \cite{Gabriel2020ArtificialIntelligenceValuesAlignment, Sorensen2024Roadmap}. 
Work on pluralistic alignment further shows that evaluation criteria (how alignment is assessed) and relevant populations (whose judgments are considered) cannot be treated as universal \cite{Rastogi2025Whose}. Taken together, these issues point to a deeper conceptual gap: \textbf{the absence of a systematic way to specify how sociotechnical alignment defines, justifies, and evaluates socially desirable AI behavior}.

The present work addresses this gap by structuring sociotechnical alignment along four interrelated dimensions (\Cref{fig:framework}): the \textit{alignment target} specifying desired system behavior, the \textit{normative concept} defining what the system aligns to, the \textit{alignment methodology} for modeling and evaluation, and the \textit{theoretical framework} justifying these choices across contexts and populations. We ground the normative-concept dimension in established social-scientific frameworks for conceptualizing sociobehavioral desirability (\S\ref{sec:normative-concepts}). We identify three key normative concepts---\emph{values}, \emph{moral judgments}, and \emph{socially enforced norms}---and examine how they are formalized across levels of analysis, from individual-level preferences to universal normative claims. This synthesis provides a basis for applying these concepts in alignment settings and for relating them to evaluations of AI behavior across contexts and populations.

Next, we use the four dimensions to examine how sociotechnical alignment is specified in practice. To this end, we construct a corpus of 281 sociotechnical alignment papers\footnote{The corpus and the code used to build it are available at \scriptsize{\url{https://sociotechnicalalignment.github.io/}}.} from the ACL Anthology\footnote{ACL Anthology allows for large-scale analysis within a manageable scope. Extending the analysis to other venues is a direction for future work.} by filtering for publications with alignment- and normativity-related keywords, manually labeling titles and abstracts with alignment targets, and conducting targeted readings to identify, where possible, theoretical frameworks and design choices underlying the alignment methodology (\S\ref{subsec:corpus-build}). The analysis of this corpus shows that sociotechnical alignment research focuses primarily on technical aspects and safety constraints, with remaining work on bias, fairness, and personalization, and comparatively little on population-level value and moral compatibility (\S\ref{sec:results-mapping}). Moreover, normative concepts are often left unspecified or conflated with alignment targets; target populations are frequently underdefined; and design choices are rarely theoretically justified (\S\ref{subsec:results-term-imprecision} and \S\ref{sec:slippage}).

Taken together, our findings show why a four-dimensional sociotechnical alignment specification is needed: the problems we identify are not isolated omissions, but recurring disconnects between \textbf{\textit{what is being aligned, for whom, by which methods, and on what normative grounds}}. Building on this diagnosis, we derive practical recommendations for alignment research (\S\ref{sec:recommendations}). We argue for theory-grounded approaches, in which normative assumptions are explicit, consistently reflected across data, modeling, and evaluation, and appropriately scoped to represented populations. To support this, we provide a mapping of theoretical frameworks onto practical alignment decisions, linking settings such as personalization, moral conflict, and cross-cultural deployment to appropriate forms of normative grounding. These recommendations aim to support cumulative progress by making alignment approaches more comparable and reliably deployable.

\section{Related Surveys}
\paragraph{Broad alignment overviews.} \citet{Wang2024OntheEssence} situate alignment historically, trace its conceptual origins, examine its mathematical foundations, and review major alignment methods. \citet{fernandes-etal-2023-bridging} survey the use of human feedback in natural language generation, and \citet{Jiang2025ASurveyofRLHF} review preference learning for LLMs, covering feedback sources and formats, preference modeling, and evaluation. Together, these surveys capture the scope of alignment research, but they generally treat normativity at a high level. They show which alignment problems are being studied without much focus on the sociotechnical aspects.

\paragraph{Sociotechnical perspectives.} These surveys frame alignment as an ongoing process of specification, interaction, and evaluation on technical and normative goals. \citet{kirk-etal-2023-past} survey feedback learning for subjective preferences and values, reflecting on prior work and future directions. \citet{terry2024interactiveaialignmentspecification} conceptualize interactive alignment across task specification, process, and evaluation, arguing that alignment should be understood across the full human--AI interaction loop rather than only through model optimization. \citet{lu2025alignmentsafetylargelanguage} review practical safety mechanisms and evaluation challenges for deployed LLMs, highlighting open problems in oversight, value pluralism, robustness, and continuous alignment. \citet{Ji2025AIAlignment} propose a forward and backward view of alignment: systems are aligned forward through training and backward through governance, organized around four principles: Robustness, Interpretability, Controllability, and Ethicality (RICE).  \citet{sharma2025rlhfcomprehensivesurveycultural} focus on emerging issues such as multimodal alignment, cultural fairness, and low-latency optimization. However, while these works broaden alignment beyond purely technical optimization, they still leave a gap in explaining how research concretely specifies and justifies the social desirability of AI behavior.

\paragraph{Position of this survey.}
This survey complements prior overviews by focusing on how socially desirable AI behavior is specified in practice. Whereas existing surveys synthesize alignment methods, feedback mechanisms, training paradigms, and system-level challenges, we examine how sociotechnical alignment research specifies alignment and justifies its choices. We therefore treat sociotechnical alignment not only as a problem of optimization, but also as one of specifying, grounding, and evaluating social desirability in NLP.

\label{sec:related-surveys}

\section{Normative Concepts of Sociobehavioral Desirability}
Social-scientific work generally treats \emph{sociobehavioral desirability} through related but distinct concepts---\emph{values}, \emph{moral evaluations}, and \emph{social norms}---whose meanings and effects vary by context. \textit{Values} (e.g., equality) are relatively abstract, transsituational goals or priorities that orient preferences and evaluation across situations \cite{Parsons1951SocialSystem,schwartz1992universals,schwartz2012refining,Ponizovskiyetal2019Social}. \textit{Norms} (e.g., disclosing sources) are context- and role-specific expectations about appropriate conduct, maintained through socialization, coordination, and sanctioning \cite{Durkheim1895Rules,Goffman1959Presentation,Bicchieri2006TheGrammar}. \textit{Morals} (e.g., lying is wrong) are judgments about right and wrong that carry stronger obligation and boundary-drawing force than many everyday norms, thereby legitimizing praise, blame, and social inclusion or exclusion \cite{Durkheim1912Elementary,Haidt2001Emotional,haidt2012righteous}. Across these levels, values articulate broad \emph{ends and priorities}, norms specify \textit{appropriate means} of conduct in particular settings, and morals supply \emph{evaluative standards} that render some expectations as right or wrong  \cite{Durkheim1912Elementary,Haidt2001Emotional}.
\begin{table}[t]
\centering
\resizebox{\linewidth}{!}{
\begin{tabular}{@{}p{3.6cm} p{6.7cm}@{}}
\toprule
\textbf{Level of analysis} & \textbf{Frameworks} \\
\midrule
Individual (micro) &
Rokeach's Value Survey (RVS); Social Value Orientation (SVO); Kohlberg's Theory of Moral Development (TMD) \\
\addlinespace
Individual $\leftrightarrow$ Universal &
Schwartz's Theory of Basic Human Values (TBHV); Moral Foundations Theory (MFT) \\
\addlinespace
Cultural (macro) &
Hofstede's Cultural Dimensions Theory (CDT) \\
\addlinespace
Cultural $\leftrightarrow$ Universal &
Inglehart's Post-Materialist Thesis (PMT); World Values Survey (WVS) (incl.\ Inglehart--Welzel dimensions); Human Development Sequence (HDS) \\
\addlinespace
Universal (normative) &
Gert's Common Morality \\
\addlinespace
Individual $\leftrightarrow$ Cultural (bridging typology) & \citet{degrootsteg2008value} (egoistic--altruistic--biospheric value orientations); \citet{Sheldon2009Comparing}\ (ecocentrism vs.\ anthropocentrism); \citet{Bicchieri2006TheGrammar}'s social norms framework; Norm Activation Model (NAM); Value--Belief--Norm (VBN) \\
\bottomrule
\end{tabular}}
\caption{Selection of widely used frameworks for conceptualizing and/or operationalizing values, norms, and morals across levels of analysis (a more complete version in Appendix \Cref{tab:continuum_visual_all}). Here, \emph{level of analysis} refers to the main social or conceptual scale a framework addresses, from individual preferences and moral judgments to cultural patterns and universal normative claims. Bridging frameworks connect more than one level.}
\label{tab:continuum_visual}
\end{table}

Social-scientific accounts further situate these concepts across different levels of social organization. At the \emph{individual level}, value hierarchies help explain preferences and action, while norms enter as expectations attached to roles and interaction settings \cite{Schwartz2011Values,Schwartz2012AnOO,Witte2020ANE,Durkheim1895Rules,Goffman1959Presentation}. Moral judgments distinguish ordinary norm violation from conduct experienced as obligatory, blameworthy, or unacceptable \cite{Haidt2001Emotional,GrahamHaidt2009Liberals,Grahametal2011Mapping}. At the \emph{cultural level}, the literature generally describes cultures less as possessing wholly different value structures but as differing in the relative emphasis placed on the same value dimensions, which in turn shapes which norms become salient and how strongly they are enforced \cite{Schwartz2006ATheory,schwartz1992universals,Stieger2016ParentchildPA,Leite2021HierarchicalCA,Vignoles2018InSO,Giorgisetal2023ThatsAll,Durkheim1895Rules,Goffman1959Presentation}. At the \emph{universal or normative level}, some traditions posit minimal moral baselines centered on harm avoidance, fairness, trust, and social cooperation, while still allowing substantial variation in how such concerns are interpreted and prioritized across groups \cite{gert2005Morality,gertculverclouser2006Bioethics,haidtjoseph2004intuitive,Graham2013MFT,schwartz1992universals,Schwartz2012AnOO}. Here, \emph{universality} concerns the assumption of broadly shared human intuitions, whereas \emph{normativity} concerns how those intuitions are defined and made socially binding by groups, cultures, or individuals.

A selective set of frameworks illustrates how these distinctions are typically conceptualized in the literature (for a more comprehensive overview, see \Cref{tab:continuum_visual}). For values, Schwartz’s Theory of Basic Human Values is widely used to model recurring value priorities and trade-offs across individuals and cultures \cite{schwartz1992universals,Schwartz2012AnOO,schwartz2012refining}. For moral judgment, Moral Foundations Theory describes recurrent dimensions of moral concern and variation in how groups weight them \cite{haidtjoseph2004intuitive,Haidt2001Emotional,Haidt2007TheNew,haidt2012righteous,Graham2013MFT}. For norms, \citet{Bicchieri2006TheGrammar}’s social norms framework is especially relevant because it explains norms in terms of empirical expectations, normative expectations, and conditional preferences sustained by anticipated approval or disapproval. In this broader social-science view, these frameworks do not collapse into a single account of alignment; rather, they mark different levels at which sociobehavioral desirability is theorized and measured.

In summary, \textbf{sociobehavioral desirability is %not a single, fixed standard but 
a layered and context-sensitive outcome of the interaction among values, norms, and moral judgments.} \emph{Values orient preferences}, \emph{norms define expected conduct \ul{in specific contexts}}, and \emph{moral judgments determine which expectations are treated as \ul{right or wrong}}. Accordingly, assessments of acceptable behavior vary across individuals and cultures, even where they appeal to broader normative concerns such as harm avoidance, fairness, trust, and social cooperation.
\label{sec:normative-concepts}

\section{Sociotechnical Alignment in NLP}
\begin{table}[t]
\centering
\small
\resizebox{1\linewidth}{!}{
\begin{tabular}{@{}p{0.5\linewidth}|p{0.5\linewidth}@{}}
\toprule
\textbf{Keywords} & \textbf{Stemmed Keywords} \\
\midrule
alignment, \hl{value}, \hl{moral}, \hl{feedback}, reinforcement, preferences, sociotechnical & align, value, moral, feedback, reinforc, prefer, sociotech\\
\bottomrule
\end{tabular}}
\caption{Keywords used to retrieve articles from ACL. Highlighted keywords were left unstemmed, and stemming as well as the keyword \emph{norm} were excluded due to many irrelevant matches.}
\label{tab:keywords-before-stemming}
\end{table}
In this section, we survey sociotechnical alignment literature in the ACL Anthology using the four-dimensional framework introduced above. We examine the \textit{alignment target} specifying (desired) system behavior, the \textit{normative concept} defining what the system aligns to, the \textit{alignment methodology} for modeling and evaluation, and the \textit{theoretical framework} justifying these choices.
To this end, we first construct a dedicated corpus from the ACL Anthology and then analyze how sociobehavioral desirability is specified, operationalized, and justified in alignment research.
\begin{figure*}[t]
    \centering
    \includegraphics[width=1\linewidth]{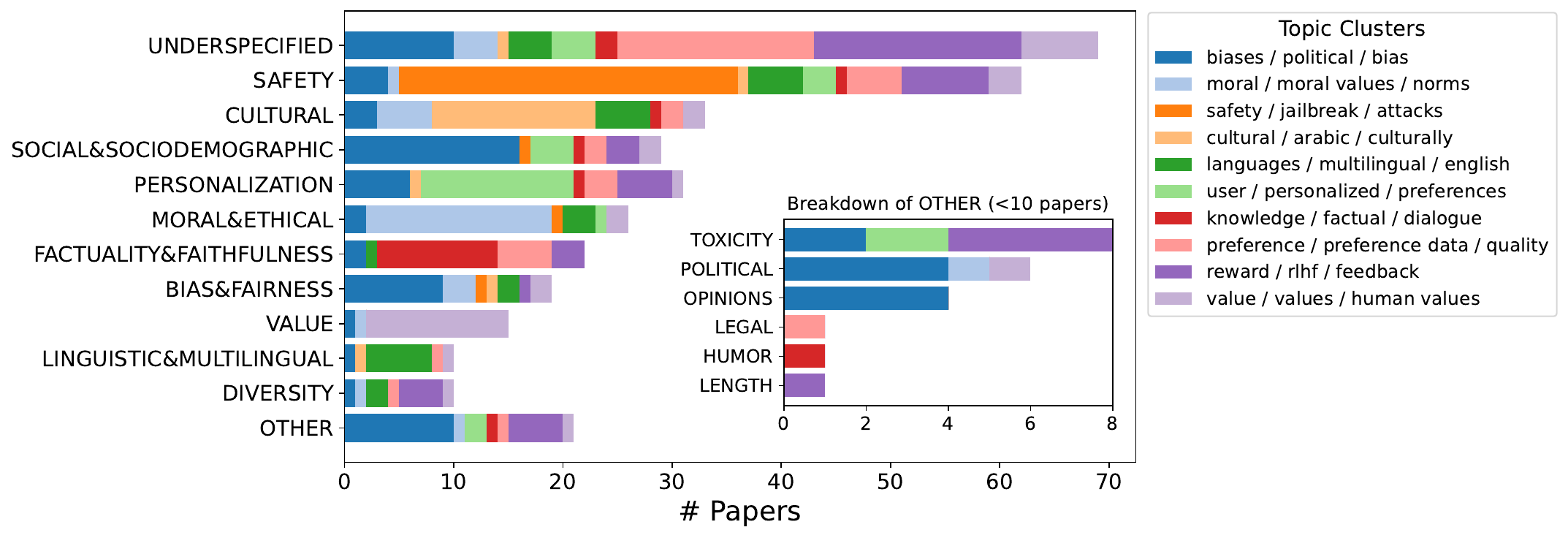}
    \caption{Distribution of annotated alignment targets across the 281-paper corpus, with automatically derived topic clusters overlaid within each manually-annotated target. Bar lengths show the number of papers per annotated target, and colored segments indicate the corresponding cluster composition from abstracts. The inset breaks down the heterogeneous OTHER category (<10 papers per subcategory).}
    \label{fig:clustered-topic-dist-bar}
\end{figure*}

\subsection{Building the Corpus}
\label{subsec:corpus-build}
We extracted all papers from ACL Anthology\footnote{\url{https://aclanthology.org/}} from 2022 to 2025\footnote{The general-purpose LLM era; cut-off: 05/08/2025.} and then applied a multi-stage filtering procedure, which we describe below.

\paragraph{Automatic filtering} We removed entries with no abstracts and filtered for abstracts in English. We then matched the remaining $37260$ paper abstracts against our keyword list in \Cref{tab:keywords-before-stemming}, retaining papers with at least two matches as a candidate set of alignment-relevant papers, resulting in a total of $1089$ papers (see \citet{kirk-etal-2023-past} for a similar survey procedure)\footnote{We matched on abstract to increase coverage. In a sample of 100 retrieved articles, we determined $\#\text{keywords}\geq2$ best balanced relevancy with the size of the retrieved set.}. The keywords can be used in different senses, therefore, we performed an additional step of automatic filtering by creating topic clusters. This resulted in $20$ clusters via topic modeling the abstracts using BERTopic with UMAP embeddings (detailed algorithms with exact parameters in Appendix \S\ref{app_sec:methods-retrieval-filtering}). We then conservatively removed all the clusters that clearly focus on topics other than alignment, such as medical NLP and code generation, which resulted in $790$ remaining papers (details in Appendix \S\ref{app_sec:methods-retrieval-filtering}). Even this reduced set remained broad and contained many false positives.

\paragraph{Manual annotation} To eliminate remaining false positives, we employed three annotators to read paper titles and abstracts (details in Appendix \S\ref{app_sec:methods-annotation}). The annotators read all $790$ titles and abstracts and identified $497$ papers with an alignment focus. They then annotated these papers based on abstracts and titles on dimensions of \textbf{(a)} sociotechnical focus, \textbf{(b)} modality, \textbf{(c)} language, and \textbf{(d)} alignment target. We retained papers with sociotechnical focus and text modality, yielding a final set of $281$ papers for analysis in our survey.

\paragraph{Assigning Alignment Targets} To capture each paper’s sociotechnical alignment target, annotators assigned free-form labels based on the abstract, producing an initial set of $25$ distinct labels (details in Appendix \S\ref{app_sec:methods-annotation}). We then merged semantically overlapping labels into broader categories to improve consistency; for example, \emph{sexism}, \emph{offensiveness}, \emph{hate}, and \emph{toxicity} were consolidated under \emph{toxicity}. Papers could receive multiple labels, and the annotators were instructed to use \emph{underspecified} when an abstract did not clearly state an alignment target. 

\paragraph{Identifying Topical Clusters} Annotation alone could not fully resolve target underspecification and terminological ambiguity. To better understand how alignment is framed in the literature, we supplemented manual annotation with topic modeling. We applied TF–IDF vectorization and Non-negative Matrix Factorization (NMF), using the scikit-learn implementations of both methods, to paper abstracts (details in Appendix \S\ref{app_sec:clustered-topics}).

Manual target labels were used to map alignment targets, while topic clusters identified broader target framings. We then conducted qualitative reading to examine how papers specified the framework's four dimensions: alignment target, normative concept, modeling or evaluation methodology, and theoretical framework. We closely read all papers in the value, moral, and cultural normative compatibility framings, where sociobehavioral desirability was most explicit. For the remaining framings, we targeted representative papers and papers exhibiting terminological ambiguity or target underspecification. We treated a design choice as theoretically justified when it was linked to a named theoretical framework, social-scientific construct, or grounded argument about target behavior or population appropriateness. This analysis supports Sections~\ref{sec:results-mapping}--\ref{sec:slippage}.

\subsection{The Sociotechnical Alignment Landscape: Alignment Targets, Normative Concepts, and Operationalization}
\label{sec:results-mapping}

Sociotechnical alignment research in NLP spans diverse targets and problems. The manually assigned target labels (y-axis of \Cref{fig:clustered-topic-dist-bar}) reveal four main patterns. \textbf{Target underspecification} is common: \emph{underspecified} is the most frequent label in the corpus. \textbf{Skewed target distribution} is also apparent: the literature focuses heavily on \emph{safety alignment}, followed by mid-sized clusters such as \emph{cultural alignment}, \emph{sociodemographic factors}, and \emph{bias and fairness}, with a long tail including \emph{humor} and \emph{verbosity}. \textbf{Values and morals are often treated as targets}: papers frequently model \emph{values} and \emph{morals} as alignment targets rather than as cross-cutting normative concepts. Finally, \textbf{terminological ambiguity} is widespread: annotators reported substantial variation in the use of key terms such as \emph{values}, \emph{safety}, and \emph{fairness}.

Further probing these patterns, our complementary topic modeling yielded ten distinct clusters shown in the legend of \Cref{fig:clustered-topic-dist-bar} and in \Cref{tab:topic_distribution}. The overlaps and divergences between these clusters and the manually assigned target labels in \Cref{fig:clustered-topic-dist-bar} indicate that the two approaches are complementary. Most automatically assigned cluster labels broadly align with the corresponding manually labeled targets. Where they diverge, the mismatches often point to deeper \emph{\textbf{conceptual underspecification}}, \emph{\textbf{terminological ambiguity}}---for example, omitted key terms or shared keywords used with different meanings across papers---or differences in what each approach captures, such as the method, the alignment target, or the normative concept emphasized. 

To better communicate how sociotechnical alignment is operationalized in the literature, we consolidate these automatically identified clusters under broader framing categories (cf. \Cref{tab:topic_distribution}): Alignment as \emph{\textbf{(a)} preference aggregation and optimization} (T1, T2), \emph{\textbf{(b)} bias and fairness assessment} (T0), \emph{\textbf{(c)} safety constraints} (T3), \emph{\textbf{(d)} personalization} (T5), \emph{\textbf{(e)} multilingual and cross-lingual transfer} (T7), \emph{\textbf{(f)} epistemic reliability} (T8), and \emph{\textbf{(g)} value and moral compatibility} (T4, T6, T9). We summarize these target framings with attention to the \emph{normative concepts} invoked, the \emph{methodologies} used to model and evaluate them, and the \emph{theoretical frameworks} used to justify these choices.

\subsubsection*{(A) Alignment as Preference Aggregation and Optimization}
A central alignment framing ($75$ papers) treats alignment as \textbf{optimizing behavior under (often multiple and conflicting) objectives} (e.g., helpfulness, harmlessness, honesty) \cite{guo-etal-2024-controllable, zhou-etal-2024-beyond_v2, zhou-etal-2024-wpo}. The works in this category are generally domain-agnostic and are shaped by the availability of \emph{preference} data, which is used as the normative concept through which social alignment is modeled. They treat \emph{``social preferences''} as heterogeneous and group-dependent signals, where the technical goal (optimization) is driven by human label variation in evaluative judgments \citep{kumar-etal-2025-compo, zeng-etal-2024-diversified}. However, the mechanisms that guide these social preferences (values, morals, and norms) are typically modeled indirectly (via preference ratings) rather than through explicit constructs or theories \citep{kumar-etal-2025-compo, nath-etal-2025-dpl}.

\subsubsection*{(B) Alignment as Public-Facing Bias and Fairness Assessment}
A prominent bias-focused cluster ($45$ papers) frames alignment as \textbf{the measurement and mitigation of biases in LMs} (e.g., demographic and political), with emphasis on how prompts elicit differential behavior across groups and how models reproduce stereotypes (e.g., racial and gendered) in generated text \citep{huang-xiong-2024-cbbq, liu-etal-2024-evaluating-large}. Rather than treating alignment solely as reward maximization, this line of work operationalizes alignment through public-facing fairness criteria---quantifying bias, comparing outcomes across demographic groups, and auditing politically charged or sensitive settings where harms manifest in model outputs \citep{nghiem-etal-2024-gotta, potter-etal-2024-hidden}. These works frame sociotechnical alignment in terms of \emph{disparities in model outputs across demographic, cultural, or political groups} under controlled prompting and evaluation settings, connecting those differences to potential equity and representational harms \citep{asad-etal-2025-beautiful, sun-etal-2025-aligned}. Although the overall alignment target is often clear, the normative basis for these socially contested objectives frequently remains implicit.
\textbf{}\begin{table}[t]
\centering
\resizebox{1\linewidth}{!}{
\begin{tabular}{@{}lr@{}} %p{0.8\linewidth}
\toprule
\textbf{Cluster: Alignment Framing (Topic Clusters)} & \textbf{\#} \\
\midrule
\textbf{0:} Bias, veracity, and stereotyping effects & 45 \\
\textbf{1:} RLHF, reward modeling, and learning from human feedback & 39 \\
\textbf{2:} Preference data quality and optimization, and response ranking & 36 \\
\textbf{3:} Jailbreaks, red-teaming, and safety against harmful behavior & 32 \\
\textbf{4:} Value theory operationalization and value datasets/principles & 28 \\
\textbf{5:} Personalization, user modeling, and persona conditioning & 25 \\
\textbf{6:} Moral belief evaluation and moral psychology benchmarks & 22 \\
\textbf{7:} Multilingual instruction tuning and cross-lingual alignment & 21 \\
\textbf{8:} Factuality, consistency, and knowledge calibration in dialogue & 17 \\
\textbf{9:} Cultural alignment and cross-cultural evaluation & 16 \\
\bottomrule
\end{tabular}}
\caption{Distribution of thematic clusters in the corpus ($N=281$). Cluster labels were derived from top TF--IDF terms in paper abstracts, and capture how the literature frames sociotechnical alignment (cf. Appendix \Cref{tab:literature}).}
\label{tab:topic_distribution}
\end{table}

\subsubsection*{(C) Alignment as Safety Constraints}
Another dominant framing ($32$ papers) defines alignment in terms of \textbf{safety constraints}, often under adversarial prompting and distribution shift \citep{xu-etal-2024-safedecoding, zhao-etal-2024-defending-large}. This includes \emph{inference-time defenses} and \emph{training-time safety specialization} \citep{liu-etal-2024-alignment, zhou-etal-2025-lssf}. These works often treat sociotechnical alignment as a \emph{system property} (robust refusal/safe completion) that must hold even when users behave strategically \citep{cao-etal-2024-defending, guo-etal-2025-mtsa}. Here, \emph{safety} broadly covers issues ranging from universal safety governance to socially contested topics such as bias and ethics. The \emph{``social''} component typically appears as \emph{normative safety standards} and \emph{policy constraints}, operationalized through harmful-prompt sets, refusal targets, and safety scoring rubrics \citep{xu-etal-2024-course, ji-etal-2025-pku}. However, these normative safety standards are often inherited from existing datasets without systematic evaluation of their validity or representativeness (e.g., whether human ethical judgments can be translated into AI behavioral standards, or whether safety criteria developed for one population remain appropriate in another population or deployment context).

\subsubsection*{(D) Alignment as Personalization and User-Centered Evaluation}
A large theme ($25$ papers) frames alignment as \textbf{fit to specific users, intents, or contexts}: \textit{user-centered benchmarks} that define ``aligned'' responses relative to user intents and multi-intent task structure, and \textit{persona/personalization mechanisms} that condition generation on user history or persona representations to improve subjective fit \citep{wang-etal-2024-user, wu-etal-2025-aligning}. These works treat alignment as \emph{interactional and situated}: model behavior should be coherent with user profiles and goals \citep{baskar-etal-2025-cper, balepur-etal-2025-whose}. The social-science connection is often implicit (user modeling, preference heterogeneity) rather than anchored in explicit theory \citep{guan-etal-2025-survey, castricato-etal-2025-persona}.

\subsubsection*{(E) Multilingual and Cross-Lingual Alignment Transfer}
Several papers ($21$ papers) investigate \textbf{alignment across languages}, frequently treating language as a proxy for community difference and measurement shift \citep{kravchenko-etal-2025-ualign, agarwal-etal-2024-ethical}. Works often discuss cross-lingual instruction tuning, transfer of safety/morality constructs, and failure modes driven by resource imbalance or annotation mismatch \citep{chen-etal-2025-instructioncp, shen-etal-2024-language}. Social science engagement here varies: some work explicitly interprets language-community differences through cultural or normative lenses, while other work treats multilinguality primarily as a technical generalization problem \citep{jinnai-2024-cross, dang-etal-2024-rlhf}.

\subsubsection*{(F) Alignment as Epistemic Reliability in Dialogue: Factuality and Consistency}
A distinct knowledge-centered cluster ($17$ papers) frames alignment as \textbf{epistemic reliability in interaction}: producing responses that remain factually grounded, internally consistent, and stable across multi-turn dialogue \citep{xue-etal-2023-improving, bi-etal-2025-context}. The normative concept in this framing is epistemic reliability, based on the implicit assumption that users want system properties such as factuality, coherence, and consistency, though this is rarely framed as social alignment. Work in this cluster focuses on reducing conversational contradictions, improving faithful expression and retrieval of factual knowledge, and aligning a model’s internal state with its responses \citep{liang-etal-2024-learning, wu-etal-2024-synchronous}. Evaluation is dialogue-centric, assessing whether responses remain factually correct across turns and cohere with earlier context and the model’s stated beliefs, rather than relying only on one-shot truthfulness scores \citep{xue-etal-2023-improving, xie-etal-2025-improving}.

\subsubsection*{(G) Alignment as Value, Moral, and Cultural Normative Compatibility}
The works discussed below frame sociotechnical alignment in terms of values, morality, and cultural norms, along with variation in values and morality across cultures, and aim to model these factors to define and evaluate desirable model behavior. In this literature, these normative concepts are sometimes modeled as alignment to the value, moral, or normative profiles of particular populations, and sometimes as properties of the model itself, expressed as a relatively stable value profile.

\paragraph{Value alignment.}
Some papers ($28$ papers; cf. T4 in \Cref{tab:topic_distribution}) study ``values'' through constructs associated with moral philosophy and social psychology to examine value adherence, tradeoffs, or priorities \citep{yao-etal-2024-value,ye-etal-2025-generative,liu-etal-2025-whats}. Values are typically operationalized through taxonomies, survey-style items, or inferred value profiles, and evaluated by comparing model outputs to human or theory-derived value distributions. Other work uses ``values'' more loosely, for example through high-dimensional representations or benchmarked national value profiles \citep{cahyawijaya-etal-2025-high,ju-etal-2025-benchmarking}.

\paragraph{Moral alignment.}
Other work ($22$ papers; cf. T6 in \Cref{tab:topic_distribution}) studies moral alignment through moral theories, questionnaires, and benchmark tasks, sometimes drawing on cognitive-developmental accounts and moral-psychology constructs to probe models' ``moral beliefs'', disagreement, and cross-lingual or cross-group variation \cite{liu-etal-2024-evaluating-moral,abdulhai-etal-2024-moral,mostafazadeh-davani-etal-2024-d3code,alvarez-nogales-araque-2024-moral}. Morality is commonly operationalized through dilemma judgments, moral-foundation labels, or acceptability ratings, and evaluated against human judgments. But some papers operationalize morality more loosely through moral dialogue, self-correction, or policy consistency \citep{sun-etal-2023-moraldial,rao-etal-2023-ethical}.

\paragraph{Cross-cultural compatibility.}
Cross-cultural work ($16$ papers; cf. T9 in \Cref{tab:topic_distribution}) sometimes draws on cultural-dimensions theory to study cultural differences and compare model behavior with human cultural distributions \citep{masoud-etal-2025-cultural,wang-etal-2024-cdeval,cao-etal-2024-bridging}. Culture is operationalized through cultural dimension surveys, cultural scenarios, or comparisons to population-level response distributions. Other work operationalizes culture through localization, cultural sensitivity, or adaptation pipelines \citep{banerjee-etal-2025-navigating,huang-etal-2024-acegpt}.

In summary, sociotechnical alignment research is concentrated on broad technical optimization and safety constraints, while much of the remaining work addresses behavioral target-focused problems such as bias, fairness, personalization, consistency, and factuality.

\subsection{Terminological Imprecision} 
\label{subsec:results-term-imprecision}
Across the literature, papers rarely specify the dimensions of sociotechnical alignment with precision. Terms such as ``values,'' ``social alignment,'' ``safety,'' or ``fairness'' often signal an alignment concern without clarifying what, concretely, is being aligned. In particular, papers often leave unclear: (i) the alignment target specifying system behavior; (ii) the normative concept invoked; (iii) how the concept is translated into modeling or evaluation practices for that behavioral target; (iv) which population is treated as relevant and why; and (v) how far claims generalize beyond the observed data. This imprecision is most visible in underspecified contributions and in work on safety, and value, moral and cultural compatibility.

\subsection{Conceptual and Theoretical Gaps}
\label{sec:slippage}
This imprecision becomes most consequential when papers explicitly invoke values, morals, or cultural norms as alignment targets. We therefore examine how these concepts are defined, operationalized, and theoretically justified.

\paragraph{Normative concept is explicitly grounded in well-known theoretical frameworks.} For \emph{values}, grounding appears in work that treats values as measurable behavioral dispositions and evaluates them using instruments adapted from the social sciences, such as surveys, inventories, and tests \citep{zhou-etal-2023-realbehavior,qiu-etal-2022-towards,do-etal-2025-aligning,jiang-etal-2025-language,liu-etal-2025-whats,choi-etal-2025-unintended,yao-etal-2025-value}. For \emph{morals}, grounding appears in work that treats morality as a measurable pattern of judgment and evaluates it using dilemma-based benchmarks, moral-psychology frameworks, standardized probing instruments, and disagreement-sensitive measures \citep{wu-deng-2025-counts,marraffini-etal-2024-greatest,guo-etal-2024-adaptable,greco-etal-2025-exploring,mostafazadeh-davani-etal-2024-d3code,falk-lapesa-2025-mining,abdulhai-etal-2024-moral,fraser-etal-2022-moral,ramezani-xu-2023-knowledge,liu-etal-2024-evaluating-moral,alvarez-nogales-araque-2024-moral}. For \emph{culture}, grounding appears in work that treats cultural norms as measurable patterns of appropriateness and variation and evaluates them using cultural-dimension frameworks, survey-conditioned dialogue settings, and culture-targeted adaptation signals \citep{xu-etal-2025-self,arora-etal-2023-probing,wang-etal-2024-cdeval,masoud-etal-2025-cultural,cao-etal-2024-bridging,liu-etal-2025-cultural}.

\paragraph{Normative concept is made explicit but definition is context- and method-specific.} Here, works point to a theoretical lineage or framework while keeping the normative concept more method- or context-specific. This includes organizational ``value'' compliance framed as alignment to explicitly stated organizational values, modeling ``values'' as patterns in a model’s internal representations, nation-grounded benchmarking framed around ``values'', psychometrics-\emph{motivated} value measurement, generative approaches that examine how language influences value expression, culture-specific ethical instruction datasets, and moral-knowledge injection framed as an inspired operationalization of ``moral values'' \citep{mittal-etal-2025-protect,cahyawijaya-etal-2025-high,ju-etal-2025-benchmarking,han-etal-2025-value,ye-etal-2025-generative,hu-etal-2024-language,hakim-etal-2025-anak,zhang-etal-2024-moka}.

\paragraph{Normative concept is loosely defined.} For \emph{values}, this underspecification appears in work where ``values'' functions as an umbrella term for heterogeneous, and at times mismatched, objectives defined primarily by what can be induced, probed, or optimized. These include natural-language value statements, implicit preference signals from human feedback, constraints defined as socially beneficial, representational directions in activation space, population-level or demographic correlates, causal value graphs, prioritized rule sets, stakeholder priorities, and context-specific safety or morality dimensions \citep{bang-etal-2023-enabling,liu-etal-2022-aligning,ammanabrolu-etal-2022-aligning,jin-etal-2025-internal,dong-etal-2025-contrans,lee-etal-2024-kornat,liu-etal-2024-generation-gap,al-ali-libovicky-2024-gender,kang-etal-2025-values,lu-etal-2024-sofa,xu-etal-2025-towards,marco-etal-2025-reader,shetty-etal-2025-vital,huang-etal-2024-flames}. For \emph{morals}, underspecification appears in work that operationalizes morality as universal right-or-wrong judgment, norm compliance, policy consistency, feedback-driven self-correction, values or lessons extracted from arguments or stories, association-based comparisons of moral values, or aggregate collective reasoning, without grounding these targets in an explicit theoretical framework \citep{leteno-etal-2025-histoiresmorales,guan-etal-2022-corpus,emelin-etal-2021-moral,sun-etal-2023-moraldial,liu-etal-2025-smaller,rao-etal-2023-ethical,paulissen-wendt-2023-lauri,hobson-etal-2024-story,xiang-etal-2025-comparing,yuan-etal-2025-probabilistic}. For \emph{culture}, underspecification appears in work that operationalizes cultural norms through sensitivity, community-grounded competence, or localization pipelines, rather than treating them as an explicitly theorized target \citep{banerjee-etal-2025-navigating,feng-etal-2025-culfit,magdy-etal-2025-jawaher,nacar-etal-2025-towards,huang-etal-2024-acegpt,hosseinbeigi-etal-2025-matina-culturally,Hu2022Lora,acquaye-etal-2024-susu,keleg-2025-llm,pandey-etal-2025-culturally}.
\label{sec:alignment-landscape}

\section{Implications and Practical Recommendations}
Sociotechnical alignment research in NLP is constrained less by a lack of methods than by a lack of conceptual precision within and across studies. Throughout the literature, alignment targets, normative concepts, methodologies, and population-specific considerations are often left implicit or treated interchangeably, producing mismatches across works. Clearer alignment research should therefore specify what is being aligned, to what, and for whom, and should ideally ground these choices in explicit justificatory frameworks. We distill these observations into practical recommendations for future alignment research.

\paragraph{Separate the behavioral target from the normative concepts.} Values, morals, norms, and preferences should not be treated as interchangeable labels and not be confused with socially desirable behavior. Likewise, safety, toxicity, fairness, misinformation, or personalization are alignment targets, not normative concepts. Work on sociotechnical alignment should state both the behavioral target being modeled and the normative concept invoked. If the normative concept is defined only in terms of general preferences, work should avoid using stronger normative terms such as \emph{human values} unless they justify that translation.

\paragraph{Ground concepts in theory.} When papers invoke values, morals, or (cultural) norms, they should anchor these concepts in an explicit theoretical measurement tradition. If preferences, annotations, or benchmark labels are used as proxies, that translation and its limits should be made clear.

\paragraph{Make the intended layer of alignment explicit across data, modeling, and evaluation.} Sociobehavioral desirability operates across multiple levels, including individual preferences, situational norms, cultural patterns, and broader normative baselines. Alignment research should specify which level it targets and ensure that data collection, optimization, and evaluation match that level. This includes stating whose judgments are represented, how disagreement is handled, and what scope of generalization is being claimed. 

We next provide illustrative guidelines for selecting theoretical frameworks to support this process. These examples are not exhaustive; the appropriate framework depends on modeling and deployment contexts across behavioral targets and populations.

\paragraph{Personalization and interaction design (user-level alignment).}

At the micro/individual level, \emph{Rokeach's Value Survey (RVS)} distinguishes \emph{terminal} values (desired end-states) from \emph{instrumental} values (preferred means), helping systems separate what users want from how they prefer to achieve it. In design, this can support targeted interactions, though it risks oversimplifying preferences or over-personalizing when values are inferred too confidently or treated as stable. \emph{Social Value Orientation (SVO)} helps model tradeoffs between self- and other-regarding outcomes, relevant for allocation, negotiation, and recommendation settings. \emph{Kohlberg's Theory of Moral Development} can guide explanations and refusals by varying justification style, such as rule-, harm-, or rights-based rationales, without changing the underlying policy.

\paragraph{Value conflict detection and robust framing (moralized topics).}
Frameworks bridging individual and more general moral structure are especially useful for handling controversial or morally loaded requests. \emph{Schwartz's Theory of Basic Human Values (TBHV)} provides a cross-context vocabulary for common tensions (openness to change vs.\ conservatism; self-transcendence vs.\ self-enhancement) and can be used to (a) surface tradeoffs explicitly and (b) parameterize interaction defaults (e.g., cautious vs.\ exploratory, privacy-first vs.\ convenience-first) in a way that is legible and empirically grounded. \emph{Moral Foundations Theory (MFT)} is well-suited to auditing the assistant's framing and tone: it helps identify when responses systematically privilege certain moral rhetorics (e.g., care/fairness) while alienating others (e.g., authority/loyalty/sanctity/liberty), which can drive perceived misalignment even when factual content is correct.

\paragraph{Cross-cultural deployment and localization (population-level alignment).}
For globally deployed systems, macro-level cultural frameworks can help anticipate differences in expectations and acceptability. \emph{Hofstede's Cultural Dimensions Theory} can guide localization of interaction norms, such as directness, authority, and uncertainty avoidance, and identify where one-size-fits-all alignment may fail. \emph{Inglehart's Post-Materialist Thesis} explains macro-level variation in priorities, such as security and stability versus autonomy and self-expression, motivating stratified evaluation designs. Because these frameworks are coarse and population-level, they should be treated as hypotheses rather than deterministic rules for individual users.

\paragraph{Bridging individual and collective goals (impact-sensitive domains).}
The \emph{De Groot \& Steg} egoistic--altruistic--biospheric typology can be practical for systems that advise on consumption, travel, energy, or sustainability-related decisions: it can structure how recommendations trade off cost/time (egoistic), social welfare (altruistic), and environmental impact (biospheric), and can be used for preference elicitation without requiring deep political or ideological inference.

\paragraph{Universal constraints and governance (normative floor).}
At the universal/normative level, \emph{Gert's Common Morality} can define a ``moral floor'' of constraints, such as avoiding serious harm, deception, and coercion, that remain stable across personalization and localization. This supports layered alignment: a universal normative floor, locale-specific interaction norms, and user-level personalization within those bounds.

\paragraph{Practical integration in alignment research.}
Across these stages, the frameworks are most actionable when used to: (a) design elicitation instruments (e.g., surveys and preference annotation) and stakeholder models (e.g., end users, impacted non-users, and communities), (b) translate ``values'' into interaction policies and constraints, (c) construct stratified evaluations that test for alignment variation across value profiles and cultures, and (d) support tradeoff documentation.
\label{sec:recommendations}

\section{Conclusion}
This work introduced a four-dimensional framework for specifying sociotechnical alignment and used it to analyze how alignment is specified in NLP research. Focusing on English ACL Anthology work enabled systematic comparison, with multilingual, cross-disciplinary, and non-ACL extensions left to future work. Our analysis revealed recurring underspecification across alignment targets, normative concepts, populations, and the scope of generalization, as well as a broader limitation: alignment research often lacks a shared basis for comparison, making claims difficult to justify and compare across studies.

The identified lack of specification has important consequences. When normative assumptions remain implicit, socially contingent judgments may be presented as universal, particularly in systems deployed across diverse contexts. This motivates the need for future work to explicitly articulate what is aligned, to what, for whom, and under which assumptions. At the same time, greater conceptual precision does not resolve which normative concepts should guide system behavior: social-scientific frameworks can clarify how values, norms, and moral judgments are defined and operationalized, but not which of them should govern system behavior. In this sense, rather than prescribing a single account of alignment, the proposed framework is intended to make these choices more transparent and comparable.
\label{sec:conclusion}

\section*{Limitations}
\paragraph{Scope limitations of keyword- and cluster-based retrieval.}
Our literature identification strategy relies on keyword queries and cluster-based filtering. While this approach supports systematic coverage at scale, it may under-represent relevant work that uses alternative terminology. We adopted this setup because it enabled high recall and, when combined with automatic and manual filtering, reduced both false negatives and false positives. As a result, some contributions, particularly those situated at disciplinary boundaries, may still have been omitted.

\paragraph{English-language restriction.}
We limited inclusion to English-language publications. This supported more consistent screening and synthesis, but it may also have introduced linguistic and regional bias by excluding studies published in other languages, including work shaped by local regulatory contexts, cultural norms, and region-specific system deployments. However, because the venues in our sample primarily publish in English, we believe this criterion remains broadly representative of the field.

\paragraph{Community and venue selection effects.}
Our review primarily reflects the emphases of the ACL community most visible in the sociotechnical alignment discourse as operationalized by our search strategy. Adjacent communities may conceptualize alignment differently, prioritize distinct alignment targets, or employ alternative methodological standards. Consequently, the themes and gaps we identify should be interpreted as contingent on this community sampling rather than exhaustive of all relevant works.

\section*{Ethical Considerations}

This paper examines how sociotechnical alignment research conceptualizes and operationalizes sociobehavioral desirability. Because concepts such as values, moral judgments, and social norms shape decisions about what AI systems should model, optimize, or evaluate, their use carries ethical and political implications. Treating these concepts as interchangeable or self-evident can obscure whose perspectives are being prioritized and which assumptions are embedded in alignment design.

A central ethical motivation of this work is to make these assumptions more visible. Underspecified alignment targets, normative bases, and evaluation criteria can make socially contingent judgments appear universal, especially when relevant populations are left implicit. This is particularly consequential for systems deployed across heterogeneous social contexts, where values, norms, and judgments of appropriate behavior may differ across groups. Our framework is intended to support greater transparency about these choices, rather than to prescribe a single correct account of sociobehavioral desirability.

We also emphasize that conceptual precision does not eliminate normative disagreement. Social-scientific frameworks can help clarify what kind of normative concept is being invoked, how it is operationalized, and for whom it is meant to apply, but they cannot determine by themselves which values, norms, or moral judgments should govern system behavior. Such decisions require explicit justification and, where appropriate, participatory and accountable forms of deliberation.

Finally, our recommendations should not be read as encouraging the uncritical modeling of any population-level preference, norm, or value. Social norms and group preferences may themselves encode exclusion, domination, or harm. For this reason, alignment design should distinguish between describing existing sociobehavioral patterns and endorsing them as appropriate targets for system behavior.

\bibliography{bib/custom, bib/anthology-1}
\bibliographystyle{acl_natbib}

\appendix

\section{Theoretical Frameworks}
We provide a comprehensive list of widely used frameworks for conceptualizing and/or operationalizing values, norms, and morals across levels of analysis in \Cref{tab:continuum_visual_all}.

\begin{table}[th!]
\centering
\resizebox{\linewidth}{!}{
\begin{tabular}{@{}p{3.6cm} p{6.6cm}@{}}
\toprule
\textbf{Level of analysis} & \textbf{Frameworks} \\
\midrule
Individual (micro) &
Rokeach's Value Survey (RVS); Social Value Orientation (SVO); Kohlberg's Theory of Moral Development (TMD); Social Domain Theory (moral--conventional--personal); Rest's Four-Component Model; Moral identity; Moral disengagement \\
\addlinespace
Individual $\leftrightarrow$ Universal &
Schwartz's Theory of Basic Human Values (TBHV); Moral Foundations Theory (MFT); Dual-process / social-intuitionist models of moral judgment (incl.\ Greene; Haidt) \\
\addlinespace
Individual $\leftrightarrow$ Norms (situational / interactional) &
Theory of Planned Behavior / Reasoned Action (TPB/TRA: subjective norms, perceived control); Cialdini's Focus Theory of Normative Conduct (descriptive vs.\ injunctive norms, salience); Bicchieri's social norms framework (empirical + normative expectations; conditional preferences) \\
\addlinespace
Cultural (macro) &
Hofstede's Cultural Dimensions Theory (CDT); World Values Survey (WVS) / Inglehart--Welzel cultural dimensions; Cultural tightness--looseness \\
\addlinespace
Cultural $\leftrightarrow$ Universal &
Inglehart's Post-Materialist Thesis / Human Development Sequence (HDS) \\
\addlinespace
Institutional / collective action (meso) &
Social norm enforcement, sanctioning, and informal social control (e.g., reputation, punishment); Ostrom's Institutional Analysis and Development (IAD) / rules-in-use \\
\addlinespace
Individual $\leftrightarrow$ Cultural (bridging typology) &
De Groot \& Steg (egoistic--altruistic--biospheric value orientations) \\
\addlinespace
Values $\rightarrow$ Personal norms $\rightarrow$ behavior (bridging) &
Norm Activation Model (NAM); Value--Belief--Norm (VBN) theory \\
\addlinespace
Universal (normative) &
Gert's Common Morality; Principlism in applied ethics/bioethics (autonomy, beneficence, nonmaleficence, justice) \\
\bottomrule
\end{tabular}}
\caption{Selection of widely used frameworks for conceptualizing and/or operationalizing values, norms, and morals across levels of analysis. Here, \emph{level of analysis} refers to the main social or conceptual scale a framework addresses, from individual preferences and moral judgments to cultural patterns and universal normative claims. Bridging frameworks connect more than one level.}
\label{tab:continuum_visual_all}
\end{table}

\section{Methods}
\subsection{Retrieval and Automatic Filtering}
\label{app_sec:methods-retrieval-filtering}
The ACL Anthology Corpus filtering is detailed in \Cref{filtering-algorithm}. This process resulted in $1{,}089$ papers. The initial topic clustering for further automatic filtering is detailed in \Cref{init-clustering-algorithm}. The resulting 20 clusters are displayed in \Cref{tab:cluster_names_counts}. After manual screening, we excluded papers in clusters 1, 2, 5, 6, 9, 13, 14, 15, 16, and 17, as these clusters primarily focused on topics other than alignment. This process eliminated $373$ entries and resulted in $790$ papers that required manual annotation.

\subsection{Manual Annotation}
\label{app_sec:methods-annotation}

\paragraph{Annotators} We employed three in-house annotators for this task. The annotators were from the international program of M.Sc. students in Computational Linguistics in their $3+$ semesters, with excellent English competency and sufficient knowledge in the focus field. They were compensated at the state-level hourly salary of research assistants with bachelors degree. 

To ensure strong inter-annotator agreement, the three annotators first labeled $100$ instances. Agreement was assessed using Cohen’s kappa in two rounds of $50$ instances each. After each round, the annotators discussed disagreements and reached a consensus on the disputed items, during which they were given feedback on the annotations and provided with additional clarifications. Following the first round, agreement improved by an overall $\Delta$ of $+0.21$, with $+0.25$ on \texttt{alignment} and $+0.16$ on \texttt{social aspect}. Given this clear improvement, we proceeded with individual annotation of the remaining instances after the second-round discussion.

\begin{table}[t]
    \centering
    \resizebox{0.6\linewidth}{!}{
\begin{tabular}{lrr}
\toprule
\textbf{keyword} & \textbf{\# in titles} & \textbf{\# in abstracts} \\
\midrule
align & 811 & 3344 \\
value & 153 & 1301 \\
moral & 59 & 120 \\
feedback & 223 & 1005 \\
reinforc & 185 & 757 \\
prefer & 259 & 1254 \\
sociotech & 0 & 4 \\
\bottomrule
\end{tabular}}
\caption{Statistics of keyword occurrences in ACL Anthology papers published between 2022 and August 5, 2025.}
\label{tab:placeholder}
\end{table}

\begin{algorithm}[t]
\footnotesize
\caption{ACL Anthology Filtering Pipeline}
\begin{algorithmic}[1]
\State Download ACL Anthology entries from \url{https://aclanthology.org/}
\State Filter for entries published in or after 2022
\Statex \hspace{\algorithmicindent} Remaining entries: $37{,}260$
\State Filter for entries containing an abstract
\Statex \hspace{\algorithmicindent} Remaining entries: $34{,}719$
\State Filter for entries with English-language abstracts
\Statex \hspace{\algorithmicindent} Remaining entries: $34{,}135$
\State Add keyword-match indicators for \texttt{align}, \texttt{value}, \texttt{moral}, \texttt{feedback}, \texttt{reinforc}, \texttt{prefer}, and \texttt{sociotech}
\State Retain entries with at least two keyword matches in the title or abstract
\Statex \hspace{\algorithmicindent} Remaining entries: $1{,}089$
\end{algorithmic}
\label{filtering-algorithm}
\end{algorithm}

\begin{algorithm*}[t]
\caption{Topic Modeling}
\tiny
\begin{algorithmic}[1]
\Function{BUILD\_TOPIC\_MODEL}{abstracts, random\_state}
    \State \textbf{Input:} abstracts, and random state.
    \State \textbf{Output:} Topic assignments, probabilities, and trained topic model.

    \State $embedding\_model \gets \textsc{SentenceTransformer}(\texttt{model\_name}=\texttt{all-MiniLM-L6-v2})$

    \State $umap\_model \gets \textsc{UMAP}(\texttt{min\_dist}=0.5,\ \texttt{metric}=\texttt{cosine},\ \texttt{n\_neighbors}=2,\texttt{n\_components}=2,$
    \Statex \hspace{1\algorithmicindent}
    $\texttt{low\_memory}=\texttt{False},\ \texttt{random\_state}=\text{random\_state})$

    \State $topic\_model \gets \textsc{BERTopic}(\texttt{umap\_model}=umap\_model,\ \texttt{embedding\_model}=embedding\_model,$
    \Statex \hspace{1\algorithmicindent} $\texttt{verbose}=\texttt{True},\ \texttt{nr\_topics}=20,\ \texttt{language}=\texttt{en},$
    \Statex \hspace{1\algorithmicindent} $\texttt{calculate\_probabilities}=\texttt{True})$
    
    \State $stop\_words \gets \textsc{EnglishStopWords}()$

    \For{each abstract $a$ in abstracts}
        \State $normalized\_abstract \gets \textsc{RemoveStopWords}(\texttt{text}=a,\ \texttt{stop\_words}=stop\_words)$
        \State \textsc{Store}(\texttt{value}=normalized\_abstract,\ \texttt{in}=sociotech\_topics)
    \EndFor

    \State $(topics,\ probabilities) \gets \textsc{FitTransform}(\texttt{model}=topic\_model,\ \texttt{documents}=normalized\_abstracts)$

    \State \Return $topics,\ probabilities,\ topic\_model$
\EndFunction
\end{algorithmic}
\label{init-clustering-algorithm}
\end{algorithm*}

\begin{algorithm*}[t]
\caption{Automatic topical clustering}
\tiny
\begin{algorithmic}[1]
\Function{TOPIC\_MODELING}{abstracts, n\_topics, n\_top\_words}
    \State \textbf{Input:} abstracts, number of topics, and number of top words.
    \State \textbf{Output:} Top terms per topic and dominant topic assignments.

    \State $normalized\_abstracts \gets \textsc{NormalizeText}(\texttt{texts}=abstracts)$

    \State $X \gets \textsc{TFIDF}(\texttt{texts}=normalized\_abstracts,\ \texttt{stop\_words}=\texttt{english},$
    \Statex \hspace{1\algorithmicindent} \texttt{ngram\_range}=(1,2),\ \texttt{min\_df}=2,\
    $\texttt{max\_df}=0.9,\ \texttt{max\_features}=4000)$

    \State $(W, H) \gets \textsc{NMF}(\texttt{matrix}=X,\ \texttt{n\_components}=n\_topics,$
    \Statex \hspace{1\algorithmicindent}
    $\texttt{init}=\texttt{nndsvda},\ \texttt{random\_state}=42,\ \texttt{max\_iter}=500)$

    \State $topic\_top\_terms \gets \textsc{TopTerms}(\texttt{matrix}=H,\ \texttt{n\_top\_words}=n\_top\_words)$

    \State $dominant\_topic \gets \textsc{ArgMax}(\texttt{matrix}=W,\ \texttt{axis}=1)$

    \State \Return $topic\_top\_terms,\ dominant\_topic$
\EndFunction
\end{algorithmic}
\label{sociotechnical-topic-cluster-algorithm}
\end{algorithm*}

\begin{table}[ht]
\centering
\resizebox{1\linewidth}{!}{
\begin{tabular}{@{}rlr@{}}
\toprule
\textbf{ID} & \textbf{Cluster Name} & \textbf{\#} \\
\midrule
-1  & Mixed / residual topics                                      & 285 \\
0   & LLM safety, jailbreaks, and defenses                         & 53  \\
\st{1}   & Medical and clinical LLM applications                        & 22  \\
\st{2}  & Vision-language / multimodal alignment and evaluation        & 69  \\
3   & Moral reasoning, ethics, and values in LLMs                 & 42  \\
4   & Retrieval, RAG, search, and ranking                          & 42  \\
\st{5}   & Agents, environments, and sequential decision-making         & 27  \\
\st{6}   & Recommender systems and user preference modeling             & 27  \\
7   & Persona-aware and interactive dialogue systems               & 33  \\
8   & Persuasion, opinion, media, and social communication         & 18  \\
\st{9}   & Code generation and coding assistants                        & 14  \\
10  & Cultural alignment and cross-cultural values                 & 22  \\
11  & Social bias, stereotypes, and fairness                       & 14  \\
12  & Linguistic structure and psycholinguistic analysis           & 12  \\
\st{13}  & Machine translation and multilingual generation              & 33  \\
\st{14}  & NLG evaluation, controllable generation, and metrics         & 24  \\
\st{15}  & Summarization and summary evaluation                         & 22  \\
\st{16}  & Reasoning, search-augmented inference, and self-improvement  & 38  \\
\st{17}  & Education, tutoring, and feedback for learning               & 22  \\
18  & Instruction tuning, fine-tuning, and alignment data         & 58  \\
19  & Preference optimization, RLHF, DPO, and reward modeling     & 211 \\
\bottomrule
\end{tabular}}
\caption{Clusters obtained via BERTopic modeling with UMAP embeddings of filtered ACL Anthology abstracts. The papers excluded after this step are displayed as \st{strikeout}.}
\label{tab:cluster_names_counts}
\end{table}

\begin{table}[t]
\centering
\resizebox{0.9\linewidth}{!}{
\begin{tabular}{lccc}
\toprule
\textbf{Round} & \textbf{Annotator Pair} & \textbf{Alignment} & \textbf{Social Aspect} \\
\midrule
\multirow{3}{*}{Round 1}
& Annotator 1 \& 2 & 0.79 & 0.76 \\
& Annotator 2 \& 3 & 0.45 & 0.44 \\
& Annotator 1 \& 3 & 0.42 & 0.52 \\
\cmidrule{2-4}
& \textbf{Mean} & \textbf{0.55} & \textbf{0.57} \\
\midrule
\multirow{3}{*}{Round 2}
& Annotator 1 \& 2 & 0.96 & 0.92 \\
& Annotator 2 \& 3 & 0.66 & 0.59 \\
& Annotator 1 \& 3 & 0.79 & 0.67 \\
\cmidrule{2-4}
& \textbf{Mean} & \textbf{0.80} & \textbf{0.73} \\
\bottomrule
\end{tabular}}
\caption{Inter-annotator agreement across two rounds for the \texttt{alignment} and \texttt{social aspect} annotations, measured with Cohen's $\kappa$. The agreement $\Delta$ from Round 1$\rightarrow$Round 2 measured at $0.21$ overall.}
\label{tab:inter-annotator-agreement}
\end{table}

\paragraph{Codebook} The annotators were given the items to annotate in individual CSV files and were given the codebook. They annotated the papers based on titles and abstracts on dimensions of \texttt{alignment} (TRUE/FALSE), \texttt{social aspect} (TRUE/FALSE), \texttt{modality} (Text, Image, etc.; multi-label possible), \texttt{language} (English, Indonesian, Multilingual, etc.; multi-label possible), \texttt{alignment target} (safety, bias, etc.; multi-label possible), and an optional field for notes (instructed to take notes if needed, especially if they found the item difficult to annotate).

\begin{figure}[t]
    \centering
    \includegraphics[width=0.5\linewidth]{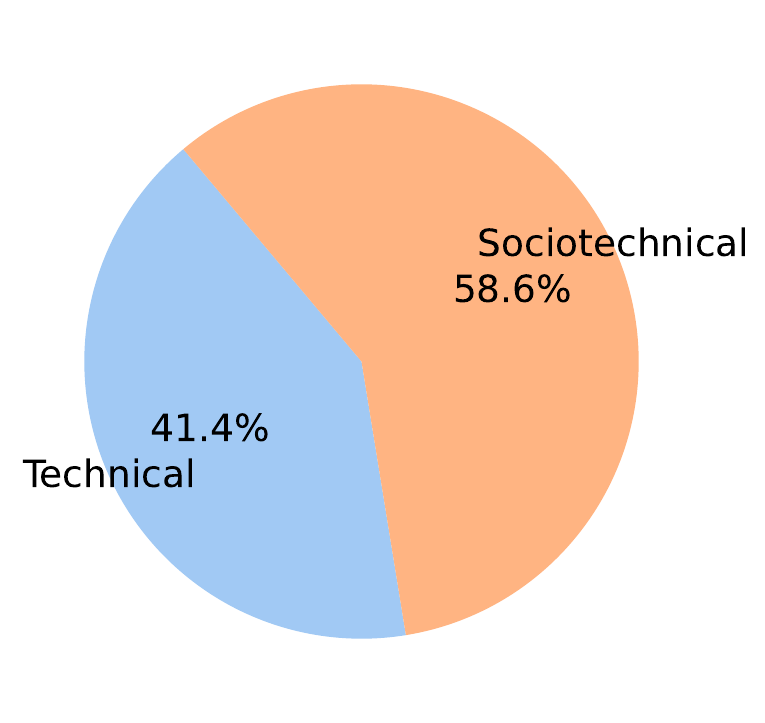}
    \caption{Distribution of purely ``technical'' vs. ``socotechnical'' alignment papers in the corpus: $206$ technical alignment and $291$ sociotechnical alignment, in total $497$ alignment focused papers.}
    \label{fig:technical-social-dist}
\end{figure}

\paragraph{Annotation Outcome} The annotators manually reviewed $790$ paper titles and abstracts, and identified $497$ alignment-focused papers, of which $291$ addressed social aspects (\Cref{fig:technical-social-dist}). Ten of these papers involved modalities beyond text only and were therefore excluded from our analysis. The final set of sociotechnical alignment papers thus comprised $281$ papers.

\subsection{Automatic topical clusters}
\label{app_sec:clustered-topics}
We detail the automatic topical clustering in \Cref{sociotechnical-topic-cluster-algorithm} (parameters are provided along with the algorithm). To obtain these clusters, we use scikit-learn’s implementation of Non-Negative Matrix Factorization (NMF) together with TF--IDF vectorization from the same library. \Cref{tab:topic_distribution} provides an overview of the clusters.

\begin{table}[]
    \centering
    \resizebox{0.9\linewidth}{!}{
    \begin{tabular}{lrl}
    \toprule
    \textbf{Assigned Label} & \textbf{\#} & \textbf{Consolidated Label} \\
    \midrule
    UNDERSPECIFIED & 69 & $\rightarrow$ UNDERSPECIFIED (69)\\
    SAFETY & 62 & $\rightarrow$ SAFETY (62)\\
    PERSONALIZATION & 31 & $\rightarrow$ PERSONALIZATION (31)\\
    CULTURAL & 30 & $\rightarrow$ CULTURAL (33)\\
    FACTUALITY & 21 & $\rightarrow$ FACTUALITY\&FAITHFULNESS (22)\\
    MORAL & 20 & $\rightarrow$ MORAL\&ETHICAL (26)\\
    BIAS & 19 & $\rightarrow$ BIAS\&FAIRNESS (19)\\
    SOCIAL & 18 & $\rightarrow$ SOCIAL\&SOCIODEMOGRAPHIC (32)\\
    VALUE & 15 & $\rightarrow$ VALUE (15)\\
    DEMOGRAPHICS & 14 & $\rightarrow$ SOCIAL\&SOCIODEMOGRAPHIC (32)\\
    DIVERSITY & 10 & $\rightarrow$ DIVERSITY (10)\\
    MULTILINGUAL & 6 & $\rightarrow$ LINGUISTICS\&MULTILINGUAL (10)\\
    ETHICAL & 6 & $\rightarrow$ MORAL\&ETHICAL (26)\\
    POLITICAL & 6 & $\rightarrow$ POLITICAL (6)\\
    TOXICITY & 6 & $\rightarrow$ TOXICITY (9)\\
    LANGUAGE & 4 & $\rightarrow$ LINGUISTICS\&MULTILINGUAL (10)\\
    OPINIONS & 4 & $\rightarrow$ OPINIONS (4)\\
    CULTURE & 3 & $\rightarrow$ CULTURAL (33)\\
    LEGAL & 1 & $\rightarrow$ LEGAL (1)\\
    HATE & 1 & $\rightarrow$ TOXICITY (9)\\
    HUMOR & 1 & $\rightarrow$ HUMOR (1)\\
    FAITHFULNESS & 1 & $\rightarrow$ FACTUALITY\&FAITHFULNESS (22)\\
    SEXISM & 1 & $\rightarrow$ TOXICITY (9)\\
    LENGTH & 1 & $\rightarrow$ LENGTH (1)\\
    OFFENSIVENESS & 1 & $\rightarrow$ TOXICITY (9)\\
    \bottomrule
    \end{tabular}}
    \caption{Annotated alignment target labels. The annotators were instructed to label papers without a clear alignment target as \emph{UNDERSPECIFIED} and were allowed to assign multiple labels to individual papers.}
    \label{tab:topic_counts_annotation}
\end{table}

\label{app:details}

\section{Literature References}
We list all the literature in our corpus in \Cref{tab:literature}.
\begin{table*}[t]
\centering
\footnotesize
\resizebox{1\linewidth}{!}{
\begin{tabular}{@{}p{0.04\linewidth}|p{0.95\linewidth}@{}}
\toprule
\textbf{ID} & \textbf{Clustered Literature} \\
\midrule
T0 & \citet{meister-etal-2025-benchmarking,murthy-etal-2025-one,galarnyk-etal-2025-inclusively,chen-etal-2025-spica,spangher-etal-2024-llms,lerner-etal-2024-whose,hu-etal-2023-decipherpref,moorjani-etal-2022-audience,huang-xiong-2024-cbbq,he-etal-2024-whose,liu-etal-2024-evaluating-large,xu-etal-2024-reasons,hsu-etal-2024-calm,malik-etal-2024-tarzan,pandita-etal-2024-rater,ma-etal-2024-potential,zubiaga-etal-2024-llm,lee-etal-2024-towards-effective,vijjini-etal-2024-socialgaze,wright-etal-2024-llm,taubenfeld-etal-2024-systematic,feng-etal-2024-modular,potter-etal-2024-hidden,nghiem-etal-2024-gotta,fulay-etal-2024-relationship,wang-etal-2024-learning-personalized,donmez-falenska-2025-understand,asad-etal-2025-beautiful,watson-etal-2025-language,zhang-etal-2025-hire,gopalakrishna-pillai-etal-2025-engagement,kirstein-etal-2025-meeting,pugachev-etal-2025-repa,rennard-etal-2025-bias,park-etal-2025-deontological,li-etal-2025-semantic-eval,chen-goldfarb-tarrant-2025-safer,zhou-di-eugenio-2025-veracity,sun-etal-2025-aligned,kim-2025-rubric,kour-etal-2025-think,ayele-etal-2024-exploring,ceron-etal-2024-beyond,kruspe-2024-musical,jones-etal-2024-multi} \\
T1 & \citet{yu-etal-2025-diverse,fu-etal-2025-unlocking,wang-etal-2025-model,kumar-etal-2025-compo,bu-etal-2025-beyond,zhang-etal-2025-metaalign,mire-etal-2025-rejected,barnhart-etal-2025-aligning,wang-etal-2024-arithmetic,liu-etal-2024-aligning,ryan-etal-2024-unintended,khatun-brown-2023-reliability,dong-etal-2023-steerlm,kirk-etal-2023-past,rosset-etal-2023-axiomatic,zhu-etal-2024-lire,park-etal-2024-disentangling,hu-etal-2024-teaching,zhou-etal-2024-beyond,hong-etal-2024-cyclealign,yao-etal-2024-pure,zeng-etal-2024-diversified,lu-etal-2024-eliminating,guo-etal-2024-controllable,chen-etal-2024-accuracy,cao-etal-2024-enhancing,kwon-etal-2024-gdpo,xie-etal-2024-calibrating,lloret-etal-2024-towards,song-etal-2025-well,xu-etal-2025-mwpo,wang-etal-2025-adversarial,riahi-samani-etal-2025-large,li-etal-2025-gradient,huang-etal-2025-deal,xie-etal-2025-bone,hamilton-2024-detecting,xia-etal-2024-aligning,zhan-etal-2024-removing} \\
T2 & \citet{roberts-etal-2025-large,hardarson-etal-2025-aligning,nath-etal-2025-dpl,zhang-etal-2025-verifiable,wu-etal-2025-pa,lee-han-2025-sentimatic,wadhwa-etal-2025-northeastern,ye-ng-2024-preference,padhi-etal-2024-value,huang-etal-2023-learning-preference,lu-etal-2023-towards,naseem-etal-2024-grounded,cheng-etal-2024-adversarial,nguyen-etal-2024-multi,amini-etal-2024-direct,fan-etal-2024-reformatted,duan-etal-2024-negating,wu-etal-2024-reuse,zhou-etal-2024-wpo,xu-etal-2024-bpo,zhang-etal-2025-persona,xiang-etal-2025-self,xu-etal-2025-cof,zhang-etal-2025-disentangling,takayama-etal-2025-persona,alnumay-etal-2025-command,nath-etal-2025-frictional,wen-etal-2025-cheems,guo-etal-2025-critiq,corrado-etal-2025-automixalign,wang-etal-2025-popalign,wang-etal-2025-a3,qiyuan-etal-2025-efficient,thorne-etal-2024-increasing,yue-etal-2024-evidence,shi-etal-2024-safer} \\
T3 & \citet{yang-etal-2025-seqar,biancotti-etal-2025-chat,jiang-etal-2025-optimizable,hazra-etal-2024-safety,wichers-etal-2024-gradient,wang-etal-2024-rlhfpoison,zhang-etal-2024-jailbreak,xu-etal-2024-safedecoding,cao-etal-2024-defending,xu-etal-2024-course,xu-etal-2024-comprehensive,yi-etal-2024-vulnerability,ren-etal-2024-codeattack,chehbouni-etal-2024-representational,zhao-etal-2024-defending-large,liu-etal-2024-alignment,xiao-etal-2024-distract,kumarage-etal-2025-towards,ganon-etal-2025-diesel,zhou-etal-2025-dont,zhang-etal-2025-intention,li-etal-2025-unraveling,liu-etal-2025-mixture,lee-etal-2025-small,zhao-etal-2025-mpo,guo-etal-2025-mtsa,zhou-etal-2025-lssf,ji-etal-2025-pku,rebedea-etal-2025-guardrails,luo-etal-2024-ensuring,chen-etal-2024-iteralign,cao-etal-2024-stealthy} \\
T4 & \citet{mittal-etal-2025-protect,cahyawijaya-etal-2025-high,bang-etal-2023-enabling,zhou-etal-2023-realbehavior,qiu-etal-2022-towards,ammanabrolu-etal-2022-aligning,liu-etal-2022-aligning,al-ali-libovicky-2024-gender,lu-etal-2024-sofa,lee-etal-2024-kornat,liu-etal-2024-generation-gap,ju-etal-2025-benchmarking,kang-etal-2025-values,marco-etal-2025-reader,do-etal-2025-aligning,dong-etal-2025-contrans,liu-etal-2025-whats,jiang-etal-2025-language,ye-etal-2025-generative,han-etal-2025-value,shetty-etal-2025-vital,jin-etal-2025-internal,xu-etal-2025-towards,choi-etal-2025-unintended,yao-etal-2025-value,huang-etal-2024-flames,yao-etal-2024-value,hu-etal-2024-language} \\
T5 & \citet{skenderi-etal-2025-team,duan-etal-2025-guidellm,baskar-etal-2025-cper,mishra-etal-2024-able,raju-etal-2024-constructing,wang-etal-2024-intent,yu-etal-2024-popalm,li-etal-2024-pixels,chen-etal-2024-towards-tool,wang-etal-2024-user,zhang-etal-2024-holistic,guan-etal-2025-survey,ji-etal-2025-enhancing,yang-etal-2025-multi,wu-etal-2025-aligning,castricato-etal-2025-persona,karine-marlin-2025-using,qin-etal-2025-maps,balepur-etal-2025-whose,hoyle-etal-2025-proxann,wang-etal-2025-know,hao-etal-2025-evaluating,oh-etal-2025-comparison,zheng-etal-2025-greaterprompt,lin-etal-2024-towards} \\
T6 & \citet{liu-etal-2025-smaller,hakim-etal-2025-anak,wu-deng-2025-counts,leteno-etal-2025-histoiresmorales,marraffini-etal-2024-greatest,paulissen-wendt-2023-lauri,rao-etal-2023-ethical,ramezani-xu-2023-knowledge,sun-etal-2023-moraldial,fraser-etal-2022-moral,guan-etal-2022-corpus,liu-etal-2024-evaluating-moral,hobson-etal-2024-story,abdulhai-etal-2024-moral,mostafazadeh-davani-etal-2024-d3code,guo-etal-2024-adaptable,yuan-etal-2025-probabilistic,xiang-etal-2025-comparing,greco-etal-2025-exploring,falk-lapesa-2025-mining,alvarez-nogales-araque-2024-moral,zhang-etal-2024-moka} \\
T7 & \citet{kravchenko-etal-2025-ualign,chen-etal-2025-instructioncp,zosa-etal-2025-got,sharma-etal-2025-faux,corral-etal-2025-pipeline,song-etal-2025-multilingual,jinnai-2024-cross,ahmad-etal-2024-generative,nguyen-etal-2024-seallms,lai-etal-2023-okapi,espana-bonet-barron-cedeno-2022-undesired,agarwal-etal-2024-ethical,shen-etal-2024-language,xu-etal-2024-exploring-multilingual,aakanksha-etal-2024-multilingual,dang-etal-2024-rlhf,yang-etal-2025-implicit,liu-etal-2025-7,kim-kim-2025-dual,wang-etal-2024-telechat,chia-etal-2024-instructeval} \\
T8 & \citet{baek-etal-2025-researchagent,wang-etal-2025-calm,xue-etal-2023-improving,liang-etal-2024-learning,chen-etal-2024-u,das-srihari-2024-uniwiz,tao-etal-2024-trust,rozner-etal-2024-knowledge,lee-etal-2024-bapo,wu-etal-2024-synchronous,liu-etal-2024-enhancing-language,ko-etal-2024-evidence,bi-etal-2025-context,xie-etal-2025-improving,yang-etal-2025-logu,li-etal-2025-big5,du-etal-2024-zhujiu} \\
T9 & \citet{xu-etal-2025-self,banerjee-etal-2025-navigating,magdy-etal-2025-jawaher,nacar-etal-2025-towards,wang-etal-2024-cdeval,alkhamissi-etal-2024-investigating,arora-etal-2023-probing,cao-etal-2024-bridging,acquaye-etal-2024-susu,hosseinbeigi-etal-2025-matina-culturally,masoud-etal-2025-cultural,pandey-etal-2025-culturally,keleg-2025-llm,liu-etal-2025-cultural,feng-etal-2025-culfit,huang-etal-2024-acegpt} \\
\bottomrule
\end{tabular}

}
\caption{The sociotechnical alignment corpus. The references listed here correspond to the literature in \Cref{tab:topic_distribution}, which is summarized in \Cref{sec:results-mapping}.}
\label{tab:literature}
\end{table*}
\label{app:references}

\end{document}